\documentclass[preprint,amsmath,amssymb]{revtex4-1}

\usepackage{graphicx}
\usepackage{bm}

\usepackage[utf8]{inputenc}
\usepackage[T1]{fontenc}
\usepackage{mathptmx}
\usepackage{tabularx}

\begin{document}

\preprint{xx}

\title{GW electronic structure calculations of Co doped ZnO}
\author{Dennis Franke}
\author{Michael Lorke}
\author{Thomas Frauenheim}
\affiliation{Bremen Center for Computational Materials Science, University of Bremen, Am Fallturm 1, 28359 Bremen, Germany}
\author{Andreia Luisa da Rosa} 
\affiliation{Bremen Center for Computational Materials Science, University of Bremen, Am Fallturm 1, 28359 Bremen, Germany}
\affiliation{Federal University of Goi\'as, Institute of Physics, Av. Esperan\c{c}a, Campus Samambaia, Goi\^ania, Brazil}

\date{today}

\begin{abstract}
  Recently the point defect responsible for the emission of cobalt in
  doped ZnO samples has been indentified\,\cite{pssb2019}. In this
  work we further extend our investigation to other point defects in
  Co-doped ZnO. We use density-functional theory and GW calculations
  to determined the orbital-resolved band structure of cobalt doped
  zinc oxide (ZnO).  We show that mainly O-p and Co-d orbitals take
  part in the process and confirm that an oxygen interstitial nearby a
  cobalt atom is a likely defect to occur in ion beam Co-doped ZnO
  samples. We also rule out that other common point defects in ZnO can
  be responsible for the observed d-d transition. Finally, we suggest
  that defect complexes involving oxygen interstitials could be used
  to promote ferromagnetism in cobalt doped ZnO
  samples.
\end{abstract}

\maketitle

Impurities in semiconductors offer the possibility of manipulating
their electronic, magnetic and optical properties. Cobalt doped ZnO
has attracted great interest for applications in diluted magnetic
semiconductors
(DMS)\cite{Co1,GWGap1,Co2,Cobalt1,PCCP2016,Nanomaterials2017,Sarsari:13,Dalpian2013,Patterson,SciRep2017}
and consequently can provide a way to possibly use the electron spin
for quantum information devices\,\cite{DMS1,DMS2}. Furthermore, doping
of ZnO with transition metals and rare-earths were found to extend the
emission range from the intrinsic band gap to the infrared
spectrum. Consequently, Co atoms can be incorporated as optical
centers into the ZnO matrix, allowing to tune its electronic and
optical properties, making it interesting for optoelectronic devices,
for example as single-photon emitters
\cite{Ronning:10,Geburt14}. Intrinsic defects, such as vacancies or
interstitials, forming defect complexes with the substitutional
impurity can change the ionization state and coordination number of
cobalt, which strongly affect the luminescence
properties\cite{Cobalt3,Lumineszenz1}.

Recent experiments have shown that after Co implantation, luminescence
signatures at 1.74–1.88 eV have been
observed\,\cite{Luminescence3,Luminescence2,pssb2019}. Co is found to
be incorporated at a zinc site leading to a 2+ oxidation
state. Luminescence spectra of ZnO nanowires ensembles with nominal Co
concentrations ranging from 0.05 to 8.0 at\% after annealing in air at
temperatures of 500-900$^{\rm o}$C shows luminescence in the visible
region assigned to cobalt incorporation. The mechanism underlying this
process has been recently identified by some of us\,\cite{pssb2019} in
which a defect involving a cobalt atom substituting a zinc atom nearby
with an oxygen interstitial atom.  In this work by employing
state-of-the-art first principles calculations, we further explore the
mechanism for this luminescene and extend our investigation to other
defects to show that other common point defects can be rule out in
participating in the observed d-d transition.

We employ density-functional theory \cite{Hohenberg1964,Kohn1965} and
many-body GW methods \cite{Hedin65} as implemented in the Vienna ab
initio simulation package (VASP) \cite{Kresse:99} to investigate the
electronic structure of Co-doped ZnO in the presence of intrinsic
defects.  The supercell used in the calculations consisted of 72 atoms
with a Co concentration of 2.7\%.  The projected augmented wave method
(PAW) has been used \cite{Kresse:99,Bloechl} to relax the structures
with the Perdew-Burke-Ernzenhof (PBE) form for the
exchange-correlation functional\,\cite{Perdew:96}. A plane wave basis
set with an energy cutoff of $E_{\text{cut}}=400\,\text{eV}$ and a
$(3\times2\times2)$ Monkhorst-Pack {\bf k}-point sampling was employed
to integrate the charge density. GW calculations have been performed
to determine the electronic structure of the doped systems. This
approach has been succesfully used in our previous
works\cite{Geburt14,Lorke:16,pssba2019,pssb2019}.

\begin{figure}[ht!]
  \centering
  \begin{tabular}{cc}
    \includegraphics[width=0.5\columnwidth,clip,]{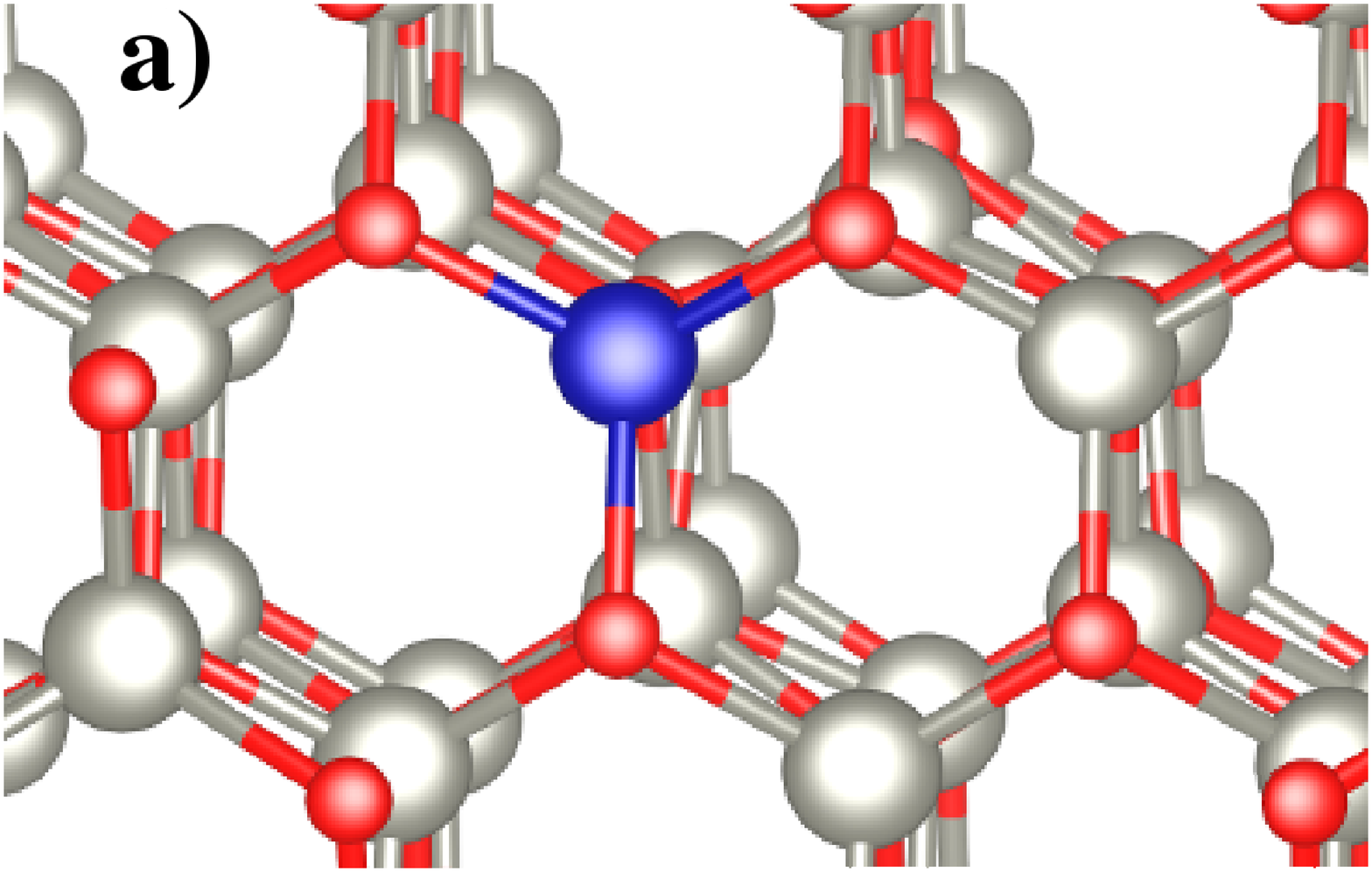}&
     \includegraphics[width=0.5\columnwidth,clip]{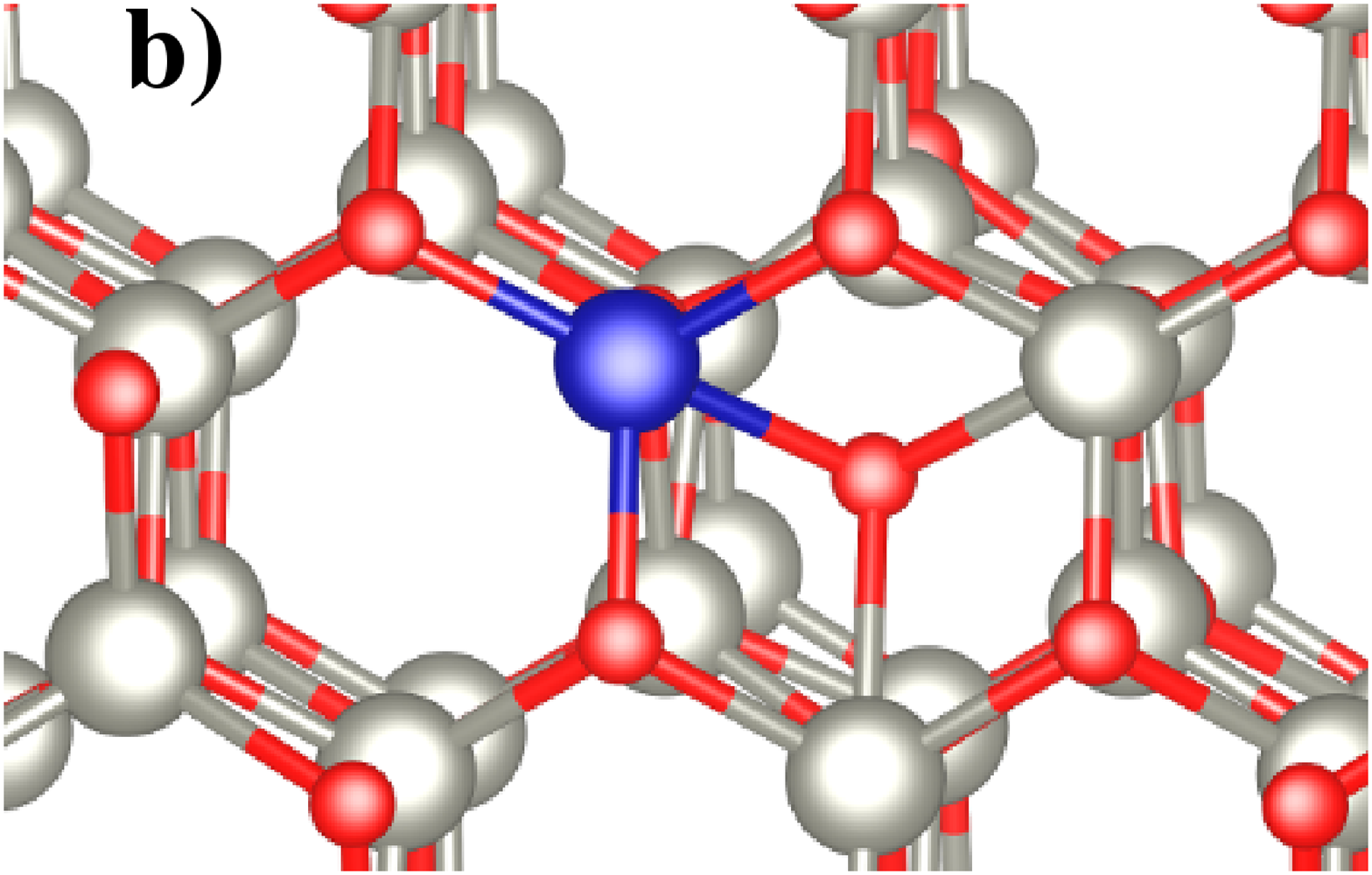} 
    \end{tabular}
  \caption{Atomic structure around the a) $\text{Co}_{\text{Zn}}$ and b) ${\rm Co_{Zn}+O_{int}}$  defects calculated within the PBE functional. Grey, red and blue spheres represent Zn, O, and Co atoms, respectively.}
\label{GEO}
\end{figure}

Since the preparation conditions can be varied by providing a O-rich
or Zn-rich environment, this can lead to drastic changes in the
electronic structure of pure
ZnO\,\cite{Lany_2010,Ronning:10,Wang:11,pssb2019}. Consequently, it is
expected that intrinsic defects would also influence the electronic
properties of doped samples. In Fig. \ref{GEO} the relaxed geometries
of (a) $\text{Co}_{\text{Zn}}$ (Co substitutional at a zinc site) is
shown.  Incorporating cobalt into the ZnO matrix at a substitutional
sites does not cause significant strain in the ZnO lattice. The cobalt
distances to nearest neighbour oxygen atoms are 1.97 {\AA} and 1.98
{\AA} for in-plane and c-direction, respectively. The Co-Zn distances
are 3.29 {\AA} for atoms in the basal plane and 3.23 {\AA} along the
$c$-direction. Fig. \ref{GEO} (b) shows the relaxed geometry of a Co
substitutional at a zinc site plus an oxygen interstitial nearby
${\rm Co_{Zn}+O_{int}}$. The Co-O distances to nearest neighbour
oxygen atoms in the ZnO lattice are 1.92 {\AA} along the c and 1.94
{\AA} for the in-plane direction. The distance to the oxygen
interstitial is 1.73 {\AA}. Although the overall change in the
structural properties of ZnO is not significant, the change in the
electronic structure of both systems compared to bare ZnO is clearly
visible, as we discuss below.

\begin{figure}[ht!]
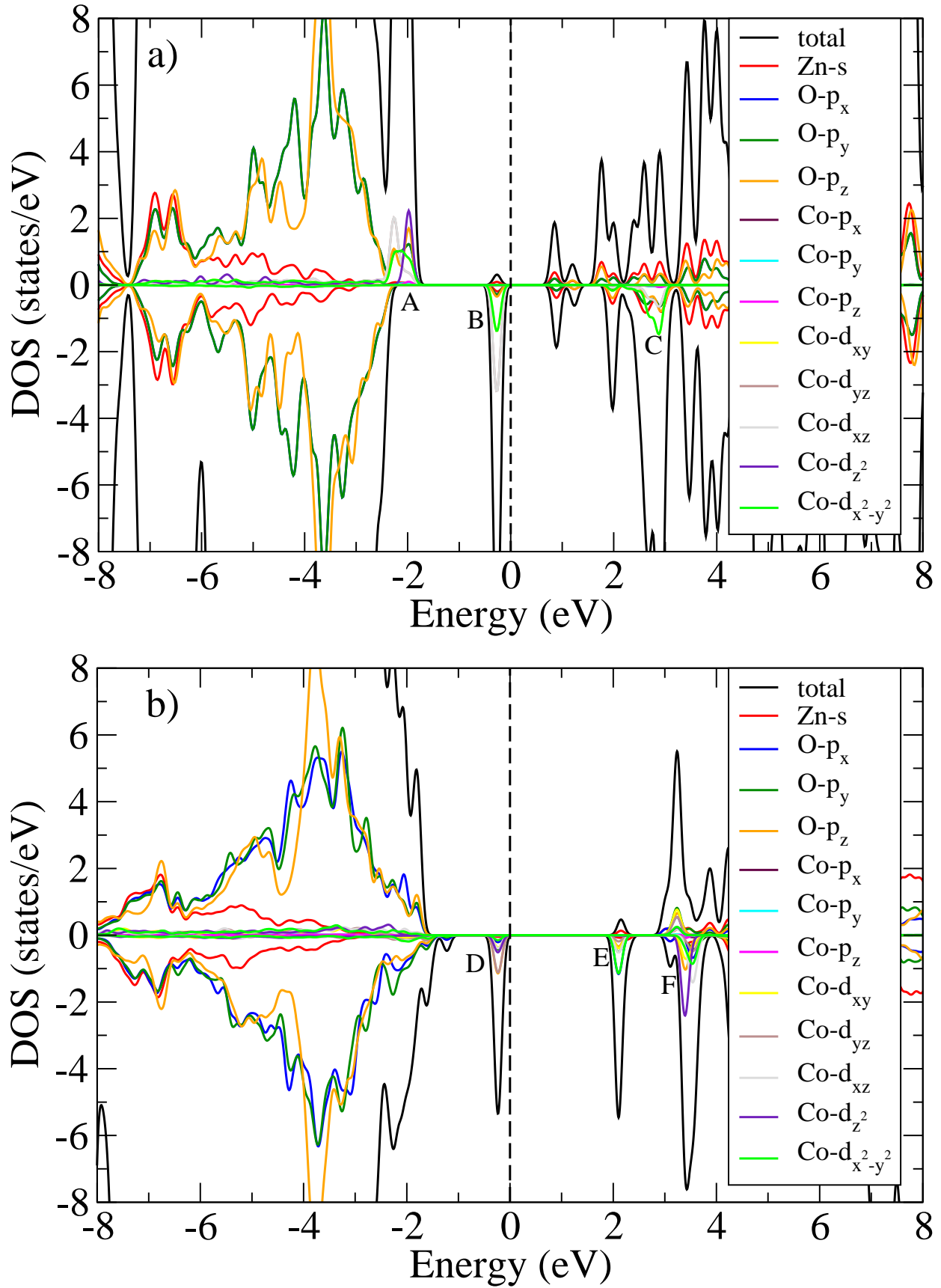

  \begin{tabular}{c}
 \includegraphics[width=1\columnwidth,clip]{DosCoSubstOrbitale.eps}\\
 \includegraphics[width=1\columnwidth,clip]{DosCoSubstOintOrbitale.eps}
\end{tabular}
\caption{Orbital projected density of states calculated  within the PBE+$\text{GW}_{0}$ approximation for a) 
$\text{Co}_{\text{Zn}}$: A denotes states close to VBM, B denotes states close to the Fermi level and C denotes states close to the CBM and for b) ${\rm Co_{Zn}+O_{int}}$: D are states close to the Fermi level, E are states close to CBM and F are higher states in the conduction band. The vertical line denotes the highest occupied state. Positive (negative) values of the DOS denote spin up (down).}
\label{DOS_orbitals}
\end{figure}

The orbital decomposed electronic structure of $\text{Co}_{\text{Zn}}$ and ${\rm
  Co_{Zn}+O_{int}}$ defects is shown in Fig.~\ref{DOS_orbitals}. We
first discuss the density-of-states of $\text{Co}_{\text{Zn}}$ shown in Fig. ~\ref{DOS_orbitals}(a). 
As Co ions occupy the Zn site in wurtzite ZnO, the ZnO
octahedral crystal field splits the Co-$3d$ states into lower $e$ and
higher t$_2$ levels. In the absence of other point defects, the
majority-spin $e$ and t$_2$ as well as the minority-spin $e$ states
are filled, while the minority spin down t$_2$ states are empty.

\begin{figure}[b!]
 \begin{tabular}{ccc}
\includegraphics[width=0.3\columnwidth,clip]{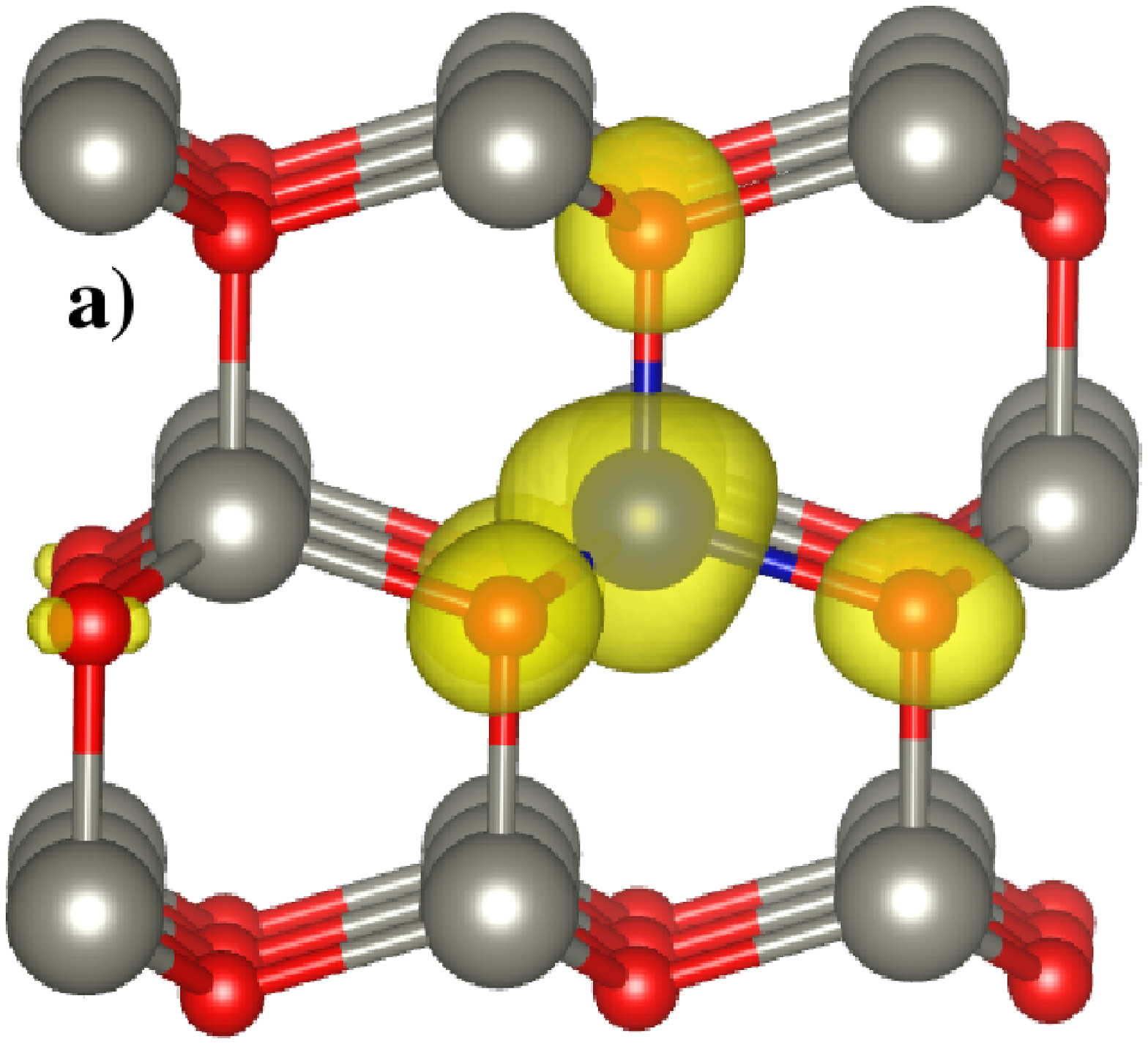}&
\includegraphics[width=0.3\columnwidth,clip]{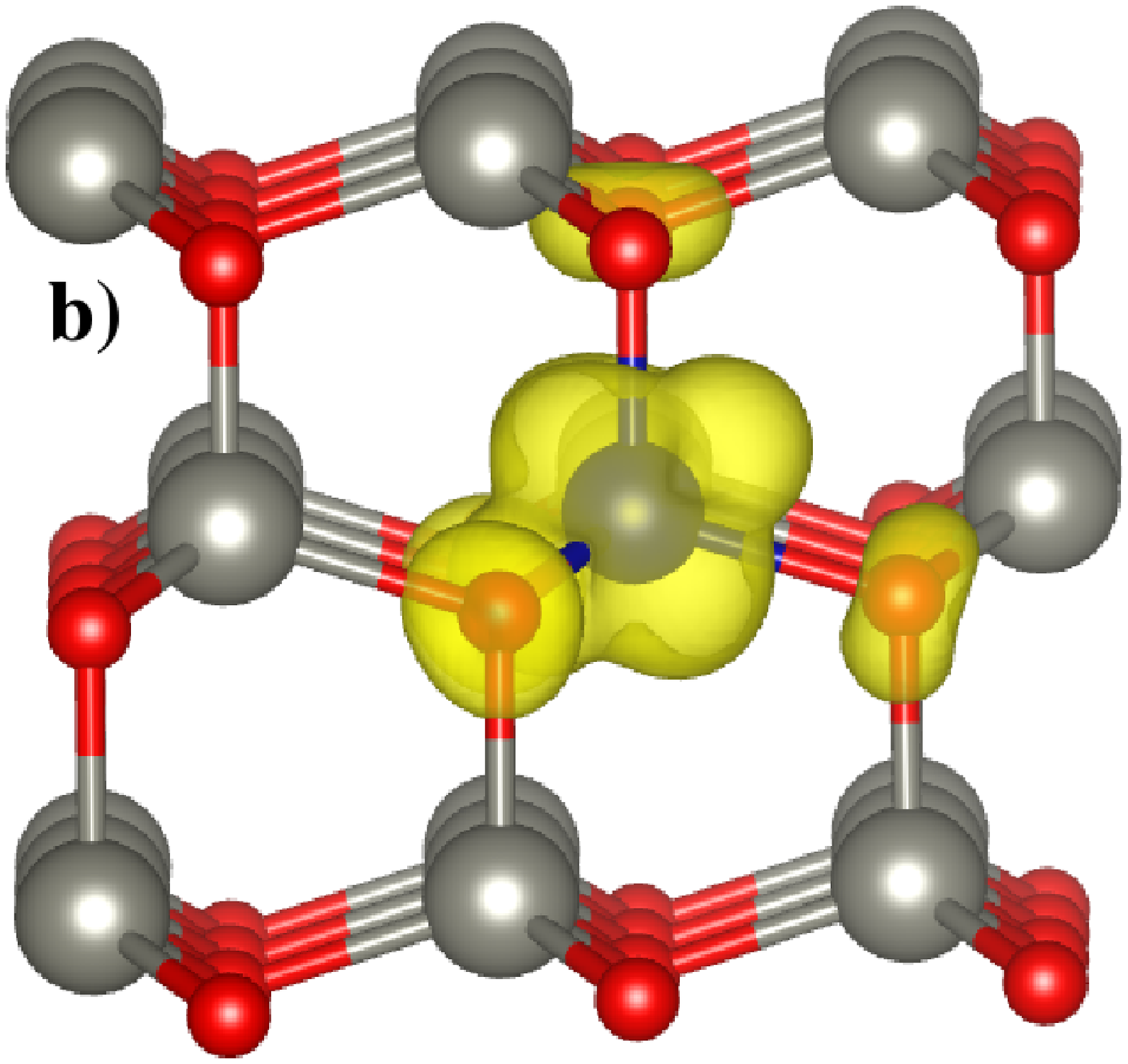}&
\includegraphics[width=0.3\columnwidth,clip]{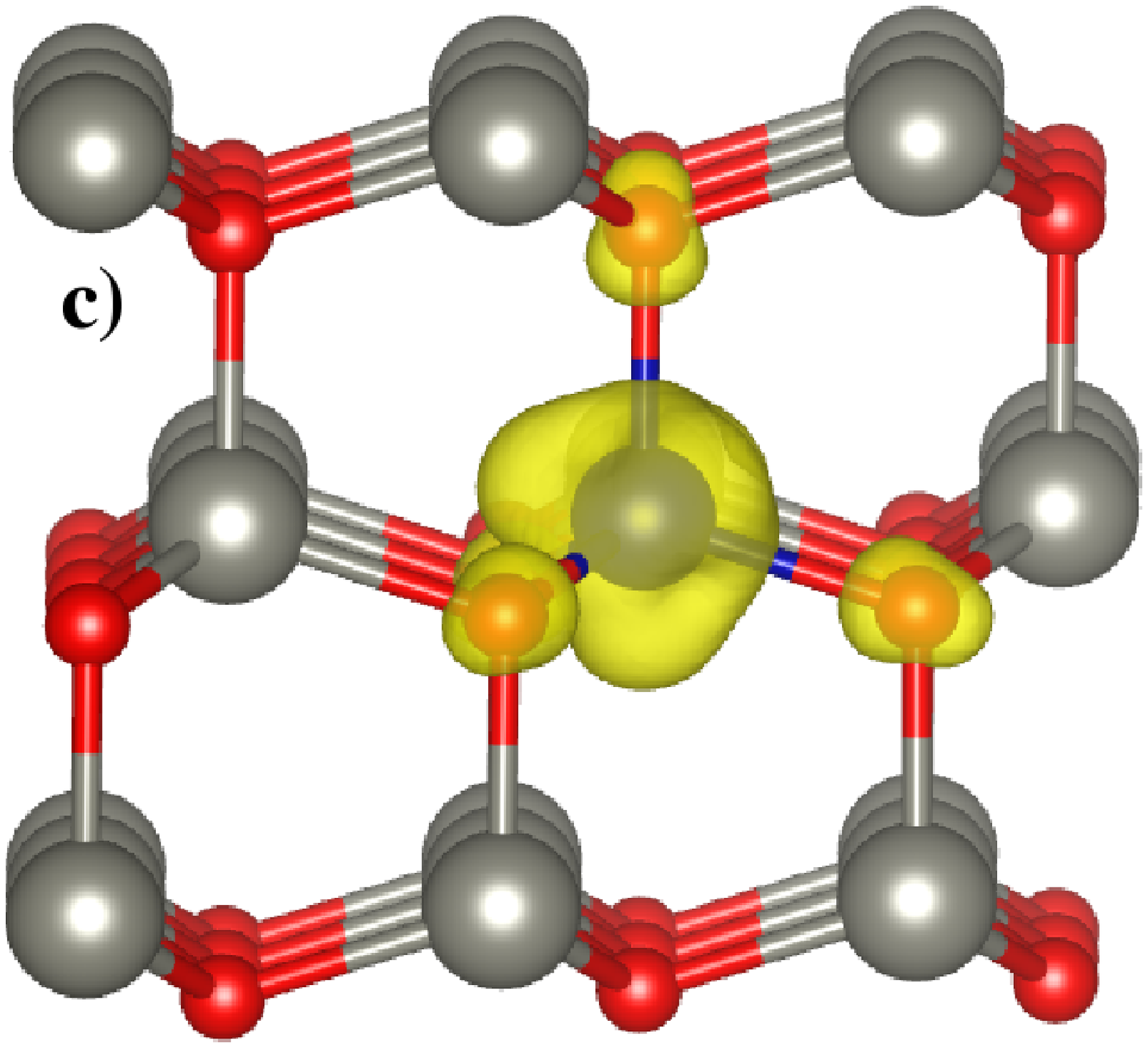}\\
\includegraphics[width=0.3\columnwidth,clip]{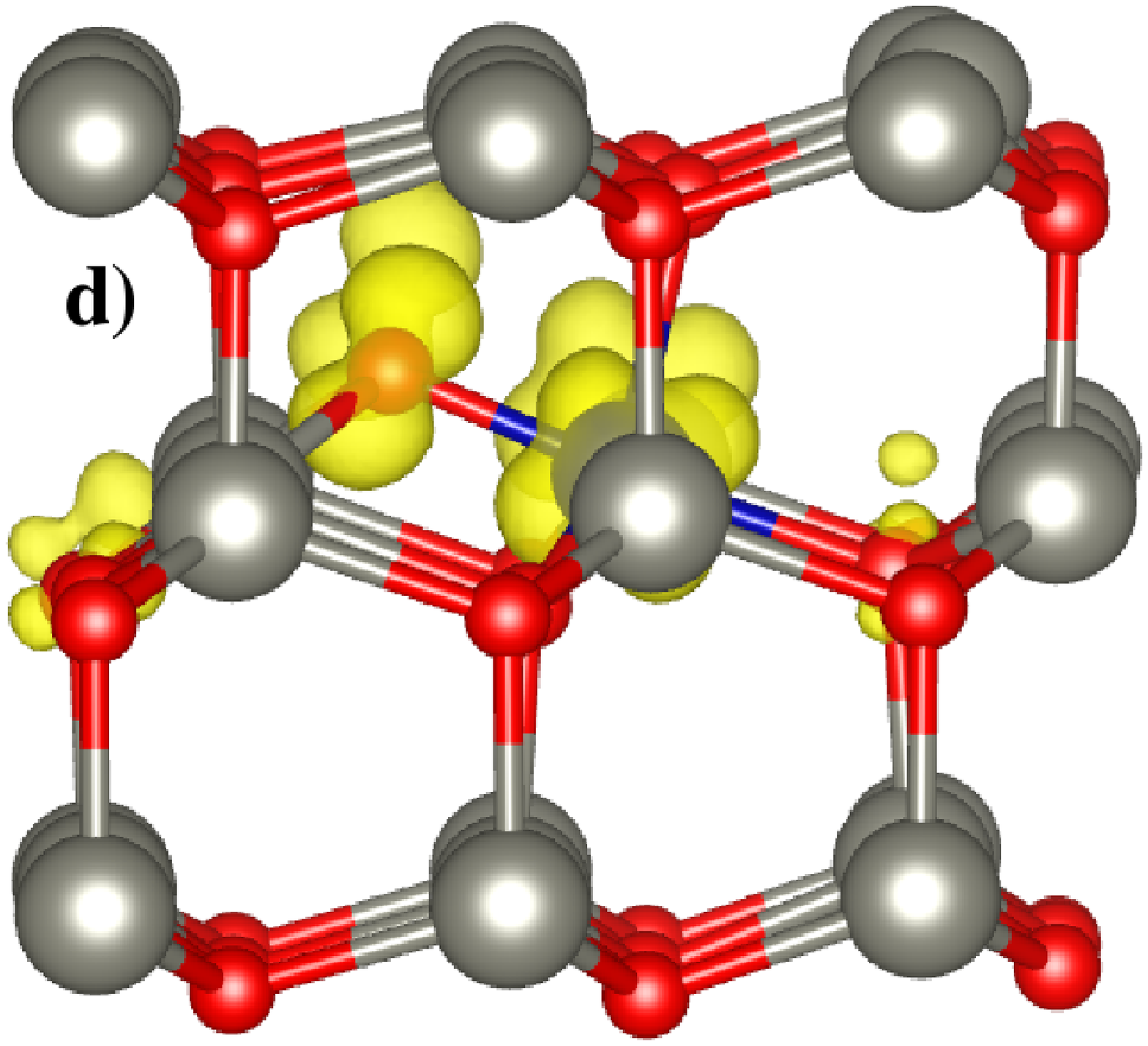}&
\includegraphics[width=0.3\columnwidth,clip]{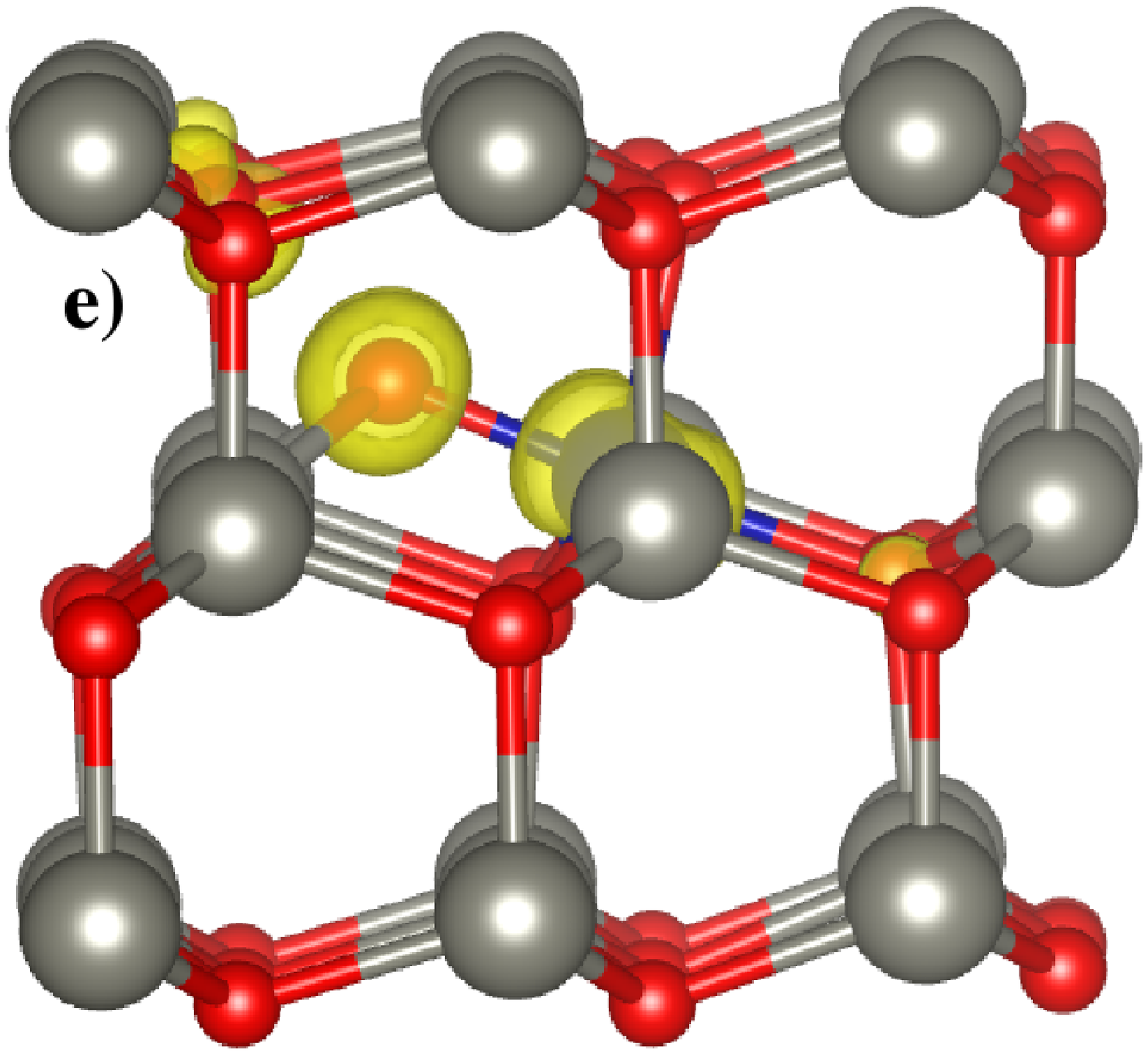}&
\includegraphics[width=0.3\columnwidth,clip]{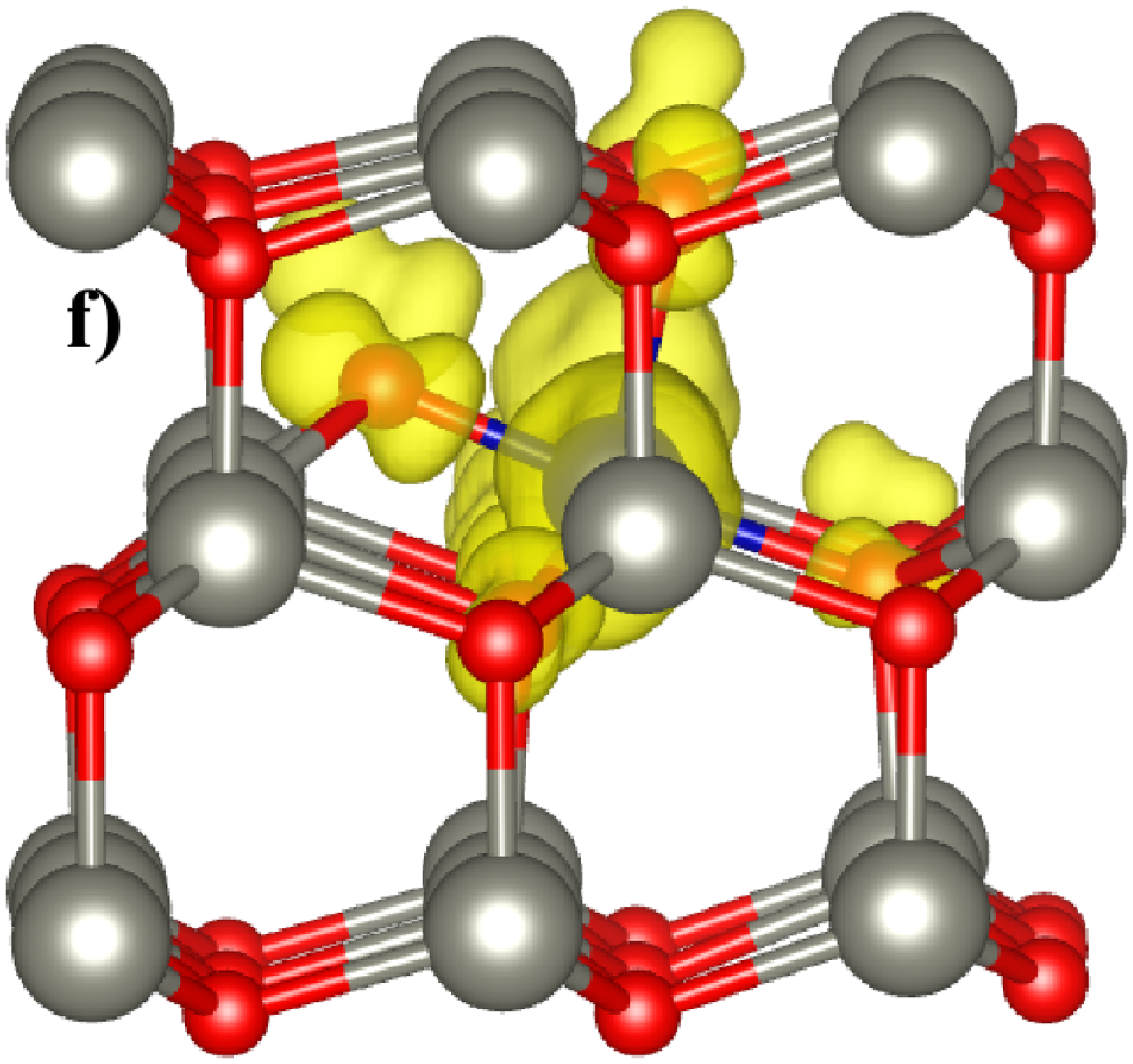}
\end{tabular}
\caption{Band projected charge density for the ${\rm Co_{Zn}}$ and ${\rm Co_{Zn}+O_{int}}$ complexes calculated within the PBE+$\text{GW}_{0}$ approximation for  ${\rm Co_{Zn}}$: a)  VBM (states located at A), b) states close to the Fermi level (states located at B), c) states located at C; and for ${\rm Co_{Zn}+O_{int}}$: d) states close to the Fermi level, e) states located at E and f) higher states in the conduction band located at F. The corresponding states A-F are shown in  Fig.\,\ref{DOS_orbitals}. Isosurface values are 0.005e/\AA$^3$.}
\label{PARCHG}
\end{figure}

For ${\rm Co_{Zn}}$ shown in Fig. \ref{DOS_orbitals} (a) we can see
that the Co-$d$ states lie inside the band gap, around 2\,eV above the
VBM. In this configuration, Co has a formal charge of 2+. The $d$
spin-up states are fully occupied, whereas the $d$ spin-down orbitals
are only partially occupied. The states labeled A close to the VBM
(around -2\,eV) are composed by O-$p$ states and Co-d states, with
larger contribution from Co-$d_{z^2}$ states. The states B, close to
Fermi level, are composed mostly by Co-$d_{x^2-y^2}$ and Co-$d_{xz}$
orbitals. The CBM (0.8-1 eV), as expected, is dominated by Zn-$4s$
states eV. Further Co-t2 states are seen inside the conduction band
around 2-3 eV. States labeled C, are composed mostly by
Co-$d_{x^2-y^2}$ and Zn-s orbitals. As a conclusion, there are no
empty d-states inside the ZnO band gap or close to the CBM to promote
the observed d-d transition. Therefore, we rule out that
$\text{Co}_{\text{Zn}}$ is responsible for the experimentally observed
luminescence reported in
Refs.\cite{Luminescence3,Luminescence2,pssb2019}.

Next we discuss the orbital projected density-of-states of
${\rm Co_{Zn}+O_{int}}$ shown in Fig.\,\ref{DOS_orbitals}(b). States
D, close to Fermi level, are composed mainly by Co-$d_{yz}$ and
O-${p_z}$. The CBM located at E (region around 2 eV) shows mostly
Co-$d_{x^2}$ and O-$p_x$ character. Higher states in the conduction
band, between 3.2 and 3.8 eV, show overlap between Co-$d_{x^2-y^2}$
and O-$p_z$ states. By comparing both $\text{Co}_{\text{Zn}}$ and
${\rm Co_{Zn}+O_{int}}$, it is clear that they possesses very
different electronic structures. The main contrast lies on the fact
that the separation between the Co low-spin states $e$ are very
different. While for ${\rm Co_{Zn}}$ the splitting between $e$ and
t$_{2}$ states is 3.2 eV, for ${\rm Co_{Zn}+O_{int}}$ it is 2.4 eV.
This means that the separation between these states is reduced in the
presence of and oxygen interstitial and therefore the overlap between
Co-d and Zn-4$s$ states is enhanced.  As reported previously, these
results are in very good agreement with the experimentally observed
luminescence signatures at 1.88 eV and 2.02
eV\,\cite{Luminescence3,Luminescence2,pssb2019}.

\begin{figure}[t]
\begin{tabular}{ccc}
\includegraphics[width=0.32\columnwidth,clip]{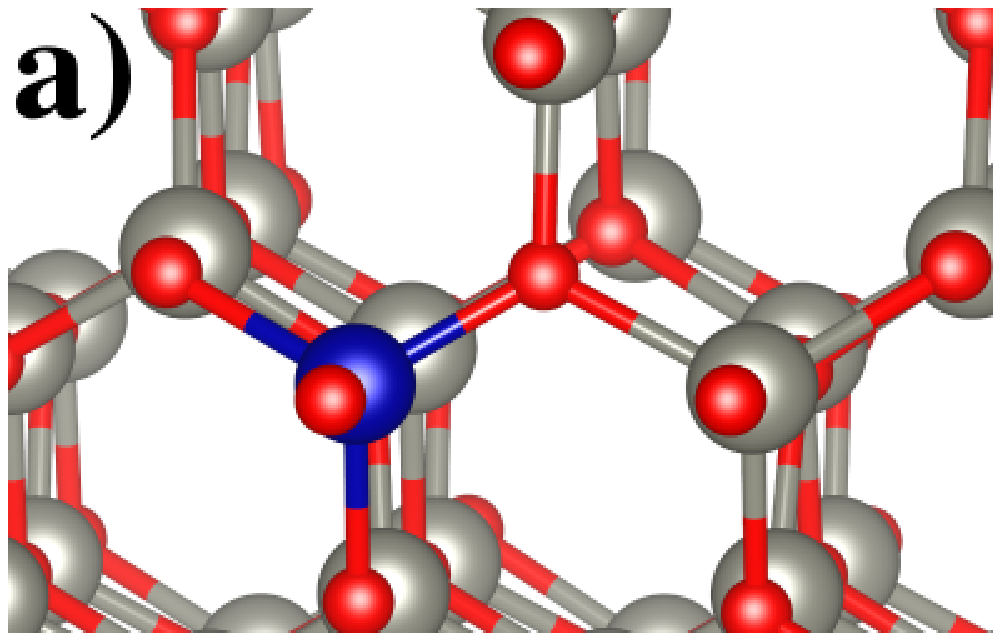}&
\includegraphics[width=0.32\columnwidth,clip]{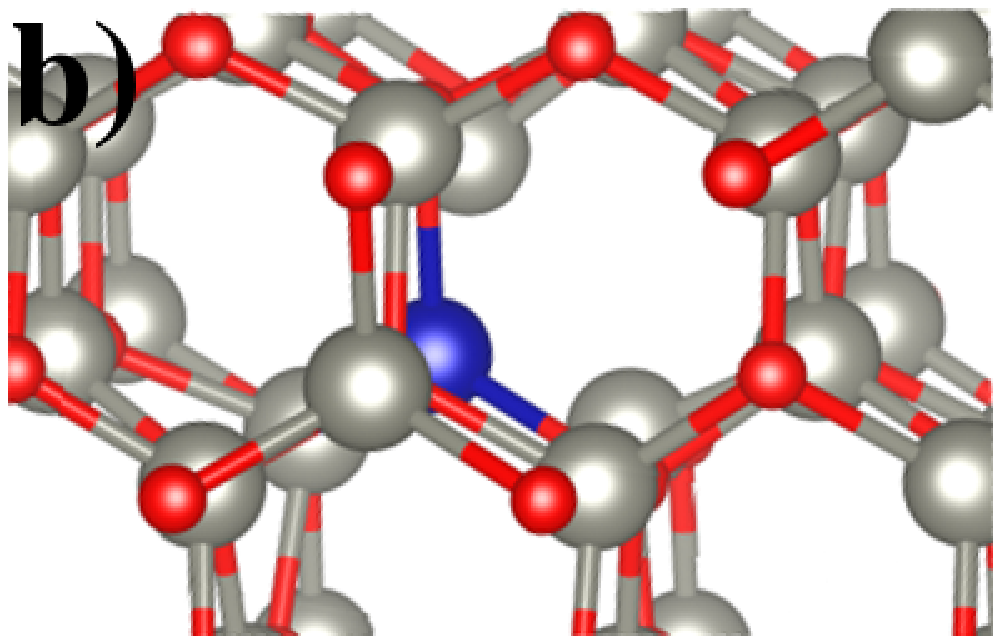}&
\includegraphics[width=0.32\columnwidth,clip]{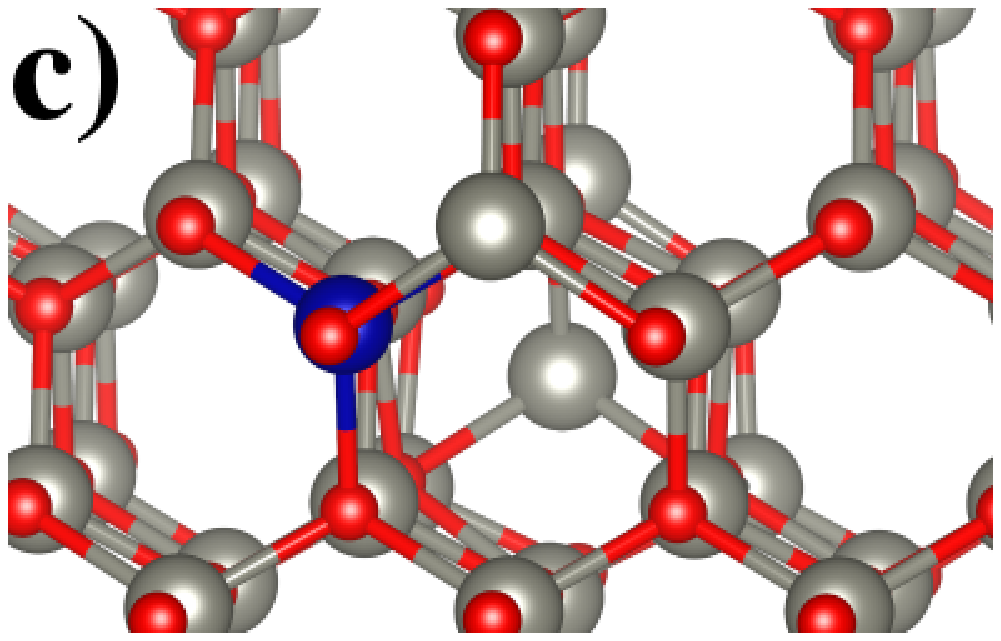}
\end{tabular}
\caption{Atomic structure around the a) ${\rm Co_{Zn}+V_{Zn}}$,
 b) ${\rm Co_{Zn}+V_O}$ and c) ${\rm Co_{Zn}+Zn_{int}}$ complexes  calculated with the PBE functional. Grey, red and blue spheres represent Zn, O, and Co atoms, respectively.}
\label{GEO_others}
\end{figure}

\begin{figure*}[ht!]
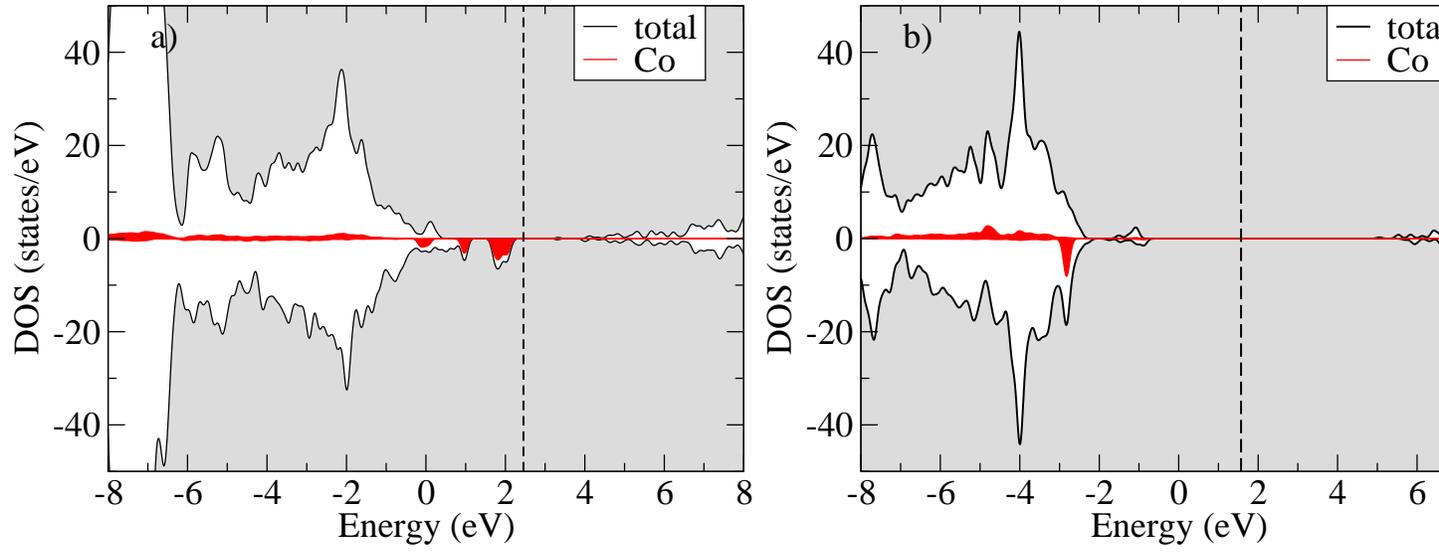

\begin{tabular}{ccc}
\includegraphics[width=0.6\columnwidth,clip]{dosCSVZN.eps}&
\includegraphics[width=0.6\columnwidth,clip]{dosCSVO.eps}&
\includegraphics[width=0.6\columnwidth,clip]{dosCSZNINT.eps}
\end{tabular}
\caption{Total and atom projected density of states for a) $\text{Co}_{\text{Zn}}+\text{V}_{\text{Zn}}$, b) $\text{Co}_{\text{Zn}}+\text{V}_{\text{O}}$ and c) $\text{Co}_{\text{Zn}}+\text{Zn}_{\text{int}}$ complexes calculated within the PBE+$\text{GW}_{0}$ approximation. The vertical line denotes the highest occupied state.  Positive (negative) values of the DOS denote spin up (down).}
\label{DOS}
\end{figure*}

In order to have further insight on the localization of these states,
the corresponding band projected charge densities calculated within
the PBE+$\text{GW}_{0}$ approximation are shown for both
${\rm Co_{Zn}}$ (Figs.\ref{DOS}(a)-(c)) and ${\rm Co_{Zn+O_{int}}}$
(Figs.\ref{PARCHG}(d)-(e)).  Fig.\ref{PARCHG}(a) shows the states
located at A, Fig.\ref{PARCHG}(b) shows the states close to the Fermi
level at B, and Fig.\ref{PARCHG}(c) shows the states located at C. All
states are very localized around the Co atom, with little or
negligible overlap with the ZnO matrix. It is evident that the states
are strongly localized on the cobalt atom with only minor
contributions on the neighboring oxygen atom for the ${\rm Co_{Zn}}$
defect. In contrast, the inclusion of the intestitial oxygen nearby
leads to a different situation, with a stronger hybridization with the
ZnO lattice.  For ${\rm Co_{Zn}+O_{int}}$ Fig.\ref{PARCHG}(d) show the
states close to the Fermi level, Fig.\ref{PARCHG}(e) shows states
located at E and Fig.\ref{PARCHG}(f) shows higher states in the
conduction band located at F. Hence in the band projected charge
densities for ${\rm Co_{Zn}+O_{int}}$ shown in \ref{PARCHG}(e) one can
cleary see that the interstitial oxygen promotes leads to additional
electrons in the band gap, making the $d-d$ transition possible.

Additionally, we have extended our investigation to Co-complexes
involving other common point defects in
ZnO\,\cite{NBANDSDep,Janotti2007,Clark:10,Lany:10}. First we discuss the relaxed atomic
structure of these complexed in ZnO. When a zinc vacancy is
close to a cobalt interstitial atom, ${\rm Co_{Zn}+V_{Zn}}$, shown in
\ref{GEO_others}(a) the Co atom slightly shifts towards the vacancy,
resulting in two different Co-Zn distances along the $c$-direction,
namely 3.21 {\AA} and 3.29 {\AA} as well in the basal plane, 3.27
{\AA} and 3.30 {\AA}. The Co-O bonds are 1.79 {\AA} and 1.90 {\AA} in
the basal plane and 1.94 {\AA} along the $c$-direction.  The next
investigated structure is an oxygen vacancy close to a cobalt
interstitial atom, ${\rm Co_{Zn}+V_O}$ shown in
\ref{GEO_others}(b). The presence of the vacancy does not disturb the
lattice strongly. The Co-O distances is 1.96 {\AA}. Co-Zn distances
are 3.27 {\AA} along the basal plane and 3.29 {\AA} along the
$c$-direction. The optimized geometry for a complex involving a zinc
interstitial and a cobalt substitutional atom complex
$\text{Co}_{\text{Zn}}+\text{Zn}_{\text{int}}$ is presented in
Fig.~\ref{GEO_others}(c).  The zinc interstital atom disturbs the
lattice and consequently the cobalt atom relaxes outwards.  The
distances between the Co interstitial and its nearest-neighbour oxygen
atoms lie in the range 1.69-2.20 {\AA}. This leads to Co-O distances
of 2.00 {\AA} in the basal plane and 2.06 {\AA} along the
$c$-direction.  The Co-$\text{Zn}_{\text{i}}$ distance is 2.21 {\AA}
and Co to second nearest neighbours is 3.25 {\AA}.

In order to verify whether any of these defects can also give rise to
luminescence in ZnO, we have calculated their corresponding total and
Co-projected density of states as shown in Fig.~\ref{DOS}.  For
${\rm Co_{Zn}+V_{Zn}}$, Fig.~\ref{DOS}(a), there are no empty Co-$d$
states inside the band gap and the formal charge of the Co atom is
close to 3+. Occupied states are located at the VBM and 1\,eV above
it. Therefore we conclude that this defect complex cannot be
responsible for the experimentally observed red emission.  The
electronic structure for the ${\rm Co_{Zn}+V_O}$ structure is
displayed in Fig.~\ref{DOS}(b).  Since there are no intra-gap states
stemming from the Co atom this defect cannot be responsible for any
luminescence observed in the experiment.  Finally, the density of
states for $\text{Co}_{\text{Zn}}+\text{Zn}_{\text{int}}$ is shown in
Fig.\,\ref{DOS}(c). Intra-gap states are now present, with empty
states located at -2.5 and -0.3 eV and occupied states are located at
0.4 and 1.2\,eV. The Co formal charge is close to 2+. However, no
clear optical transition is seen.

In order to assess the thermodynamic stability of defect complexes, we
have calculated their formation energies. Following the approach
described in Ref.\,\cite{FormationE}, the formation energy $E_{\text{f}}$ of
a neutral defect in ZnO is defined as

\begin{equation}
E_{\text{f}}=E_{\text{defect}}^{\text{tot}}-E_{\text{bulk}}^{\text{tot}}-\sum_{i}n_{i}\mu_{i},
\end{equation}

where $E_{\text{defect}}^{\text{tot}}$ is the total energy of the
respective defect complex and $E_{\text{bulk}}^{\text{tot}}$ is the
total energy of a bulk ZnO supercell. $n_{i}$ describes the number of
atoms of type $i$ that have been added or removed from the complex
with $\mu_{i}$ as the corresponding chemical potential. Since the
conditions can assume any value between Zn-rich or O-rich conditions
in a synthetization process, limits should be introduced to the
chemical potential $\mu$. In the lower limit, the material is free of
defects, whereas the upper limit corresponds to the formation of
elemental bulk/precipitation phases. This can be avoided by employing
the following condition for Co metal $\mu_{\text{Co}}\leq \mu_{\text{Co-bulk}}$.  A similar condition can be
imposed for cobalt oxide in order to avoid the formation of such
crystal phases $\mu_{\text{Co}}\leq \mu_{\text{CoO}}$, where CoO was
chosen as the upper limit. With $\mu_{\text{Zn}}$ and $\mu_{\text{O}}$ as the chemical otentials of zinc and oxygen, respectively, the ZnO chemical
potential is:

\begin{equation}
\mu_{\text{ZnO}}=\mu_{\text{Zn}}+\mu_{\text{O}}.
\end{equation}

Since the formation enthalpy of ZnO is defined as:

\begin{equation}
\Delta H^{\text{ZnO}}=E_{\text{ZnO}}^{\text{tot}}-\mu_{\text{Zn}}-\mu_{\text{O}},
\end{equation} the potential energy of oxygen can be expressed as

\begin{equation}
\mu_{\text{O}}=\mu_{\text{O}_{\text{2}}}+\lambda\Delta H^{\text{ZnO}}.
\end{equation}

Here  $\mu_{\text{O}_{\text{2}}}$ is the oxygen molecule chemical
potential and $\lambda$ is 0 (1) for oxygen rich (poor) conditions. Furthermore, we define the chemical potential of 
cobalt oxide as:

\begin{equation}
\mu_{\text{CoO}}=\Delta H^{\text{CoO}}+\mu_{\text{Co-bulk}}+\mu_{\text{O}_{\text{2}}},
\end{equation}

where $\Delta H^{\text{CoO}}$ is the formation enthalpy of cobalt oxide bulk. Finally, the chemical 
potential for cobalt can be written as:

\begin{align}
\mu_{\text{Co}}\leq \Delta H_{\text{CoO}}+\mu_{\text{Co-Bulk}}-\lambda \Delta H^{\text{ZnO}}.
\end{align}

The total energy for the respective zinc and cobalt bulk systems were
calculated using a hcp crystal structure. We find a value of
$\Delta H^{\text{ZnO}}=2.88 \,\text{eV}$ using PBE, which is in good
agreement with other GGA calculations
\cite{Enthalpy1,Enthalpy2}. Cohesive energies of
$E_{\text{c},\text{PBE}}^{\text{ZnO}}=-7.37 \, \text{eV}$ and
$E_{\text{c},\text{PBE}}^{\text{Zn}}=-1.11 \, \text{eV}$ have been
calculated, which agree very well with experimental values of
$E_{\text{c},\text{exp}}^{\text{ZnO}}=-7.52 \, \text{eV}$ \cite{CRC}
and $E_{\text{c},\text{exp}}^{\text{Zn}}=-1.35 \, \text{eV}$
\cite{Kaxiras03}, respectively.

\begin{table}
\caption{Formation energies $E_{\text{f}}$ for the intrinsic defects and defect complexes in ZnO  calculated with the PBE functional.}
\label{FormationEnergies}
\begin{ruledtabular}
\begin{tabular}{lcc}
 & \multicolumn{2}{c}{$\textrm{\ensuremath{E_{\textrm{f}}}}$ [eV]}\\
    \cline{2-3}
   defect & O-rich & O-poor\\
    \hline
    $\text{Co}_{\text{Zn}}$ & -0.91 & 0.94\\
    $\text{V}_{\text{Zn}}$ & 1.63 & 4.51\\
    $\text{V}_{\text{O}}$ & 3.74 & 0.85\\
    $\text{Zn}_{\text{i}}$ & 5.44 & 2.56\\ 
    $\text{O}_{\text{i}}$ & 4.01 & 6.89\\
    $\text{Co}_{\text{Zn}}+\text{V}_{\text{Zn}}$ & 0.25 & 4.99\\
    $\text{Co}_{\text{Zn}}+\text{V}_{\text{O}}$ & 3.41 & 2.37\\
    $\text{Co}_{\text{Zn}}+\text{Zn}_{\text{int}}$ & 14.79 & 13.76\\
    $\text{Co}_{\text{Zn}}+\text{O}_{\text{int}}$ & 1.83 & 6.57\\
\end{tabular}
\end{ruledtabular}
\end{table}

Table~\ref{FormationEnergies} displays the calculated formation
energies for the investigated defects under O-rich and O-poor
conditions. Cobalt in ZnO has a low formation energies for both
conditions in the absence of intrinsic defects (-0.91 eV, 0.94
eV). Oxygen vacancies also have a low formation energy (0.85 eV) under
O-poor conditions, whereas zinc vacancies have a higher fomation
energy (1.63 eV) under O-rich conditions. In the limit of O-rich
conditions, oxygen interstitials are also more likely to form compared
to the O-poor case. The opposite is true for zinc interstitials. These
values agree with previous works \cite{FormationenergiesWalle,JPCM2009,Lany:10}. On
the other hand, incorporating cobalt at interstitial sites is
unfavourable due to the strain and will not be discussed further
\,\cite{CoInterstitial1}.

The most stable defect complexes under O-rich (Zn-poor) conditions are
the $\text{Co}_{\text{Zn}}+\text{V}_{\text{Zn}}$ (0.25 eV) and the
$\text{Co}_{\text{Zn}}+\text{O}_{\text{int}}$ (1.83 eV) defect
complexes. So it is energetically more favourable to create a zinc
vacancy than to create an oxygen interstitial. This might be due to
the fact that because of their size, incorporating an O atom at an
interstitial site causes more stress to the ZnO matrix. This is
corroborated by the observation, that
$\text{Co}_{\text{Zn}}+\text{Zn}_{\text{int}}$ complexes have very
high formation energies under both conditions and are therefore very
unlikely to form. A possible explanation for the stability of the
$\text{Co}_{\text{Zn}}+\text{O}_{\text{int}}$ complex could be the low
diffusion barrier for oxygen interstitials
\cite{FormationenergiesWalle}. Moreover, under O-poor (Zn-rich)
conditions, only the $\text{Co}_{\text{Zn}}+\text{V}_{\text{O}}$
complex has a fairly low formation energy (2.37 eV). Interestingly,
the isolated oxygen interstitial has a fairly high formation
energy. This means that once it is formed during an experiment, it can
quickly form complexes in the material. Since oxygen octahedral
interstitials have a relatively low diffusion
barrier\cite{Janotti2007,JPCM2009}, it is likely to be produce under
O-rich conditions and to form a stable complex.

Finally, in an attempt to give some insight in ferromagnetic
properties of Co doped ZnO we briefly discuss the magnetic moments of
the isolated Co defect and the Co defect in the presence of an oxygen
interstitial.  Much has been discussed about electron mediated
ferromagnetism in transition metal doped ZnO. The energy position of
the t$^2$ minority states relative to the host conduction band is
crucial for the carrier-mediated ferromagnetism\,\cite{Sarsari:13}.  In
${\rm Co_{Zn}}$ we found an overall magnetic moment of 3$\mu_B$ per Co
atom. In this case no strong overlap between the Zn-4$s$ and Co-$d$
states is found. This means that such defect alone is unlikely to
promote ferromagnetism in ZnO. On the other hand, our results show
that an enhancement of the overlap between Zn-$4s$ CBM states and
Co-$d$ states can be achieved by tuning the growth conditions where an
excess of oxygen is provided\,\cite{pssb2019}.  This is provided by
the ${\rm Co_{Zn}+O_{int}}$ defect, where the Co minority t$_2$ states
are located close to the CBM with a strong overlap with Zn-4s
states. In this case, the total magnetic moment of the complex is 2.8
$\mu_B$, determined by a magnetic moment of 1.92 $\mu_B$ on the
cobalt, 0.2 $\mu_B$ on the oxygen interstitial atom, and smaller
contributions from the oxygen atoms surrounding the defect complex. We
propose that this scenario could favor ferromagnetism in Co doped
ZnO\,\cite{Co1,GWGap1,Co2,Cobalt1,PCCP2016,Nanomaterials2017}.

In conclusion, we have performed density-functional theory and GW
calculations to investigate the influence of neutral intrinsic defects
in Co-doped ZnO. We find that a defect involving a cobalt atom
substituting a zinc atom in the presence of an oxygen interstitial is
the most probable defect giving rise to the intra-3$d$ luminescence of
recent experimental finds\,\cite{pssb2019}. Furthermore, we suggest that
such a defect complex could help promoting ferromagnetism in
cobalt doped ZnO samples.

A. L. R. and Th. F. acknowledge funding by the DFG research group FOR
1616 “Dynamics and Interactions of Semiconductor Nanowires for
Optoelectronics”. A. L. R. also thanks the brazilian funding agencies
CNPq and FAPEG and CENAPAD for computational resources. We thank
C. Ronning, S. Geburt and M. Zapf for fruitful discussions.


\end{document}